\documentstyle[sprocl]{article}

\def\be{\begin{equation}}
\def\ee{\end{equation}}
\def\bea{\begin{eqnarray}}
\def\eea{\end{eqnarray}}

\begin{document}

\title{REACTION RATES OF NEUTRON CAPTURE BY LI-- AND BE--ISOTOPES}

\author{ H. HERNDL, R. HOFINGER, P. MOHR, H. OBERHUMMER }

\address{Institut f\"ur Kernphysik, TU Wien, Wiedner Hauptstr.~8--10,
A--1040 Wien, Austria}

\author{ T. KAJINO }

\address{Div.~Theoretical Astrophysics, Nat.~Astronomical Observatory,
Osawa 2--21--1 Mitaka, J--181 Tokyo, Japan}

\author{ N. TERASAWA }

\address{Research Institute for Physical and Chemical Research,
Hirosawa 2--1 Wako, 351--01 Saitama, Japan}

\maketitle\abstracts{Neutron capture by neutron--rich Li-- and Be--isotopes
plays a role in big--bang nucleosynthesis, especially in
its inhomogeneous version and in the $\alpha$--process occurring
in supernovae. New reaction rates for $^{7,8}$Li(n,$\gamma$)$^{8,9}$Li and
$^{9,10,11}$Be(n,$\gamma$)$^{10,11,12}$Be
have been consistently calculated using direct capture for
the nonresonant part and the Breit--Wigner formula
for the resonant part. The spectroscopic factors, spin/parity
assignments and excitation energies of the final bound and
initial resonant states have been taken from existing
experimental data whenever possible. For unstable nuclei
where this information is not experimentally available
the shell model was used to determine these quantities.}

\section{Introduction}
Evolution of the physical conditions of the universe, galaxies and stars
can be described in terms of the increase or decrease of hundreds of 
elemental abundances of atomic nuclides.  They originate from the primordial 
nucleosynthesis about fifteen billion years ago and the subsequent 
production/destruction cycle in stars and ejection into the intergalactic 
space.  It is therefore inevitable and even fundamental to study the nuclear 
processes in several astrophysical sites for a deep understanding of the 
evolution of the universe.

Cosmologically, the primordial nucleosynthesis provides a unique method 
to determine the average universal mass--density parameter $\Omega_B$.
Although the {\em homogeneous big--bang model\/} for primordial nucleosynthesis
predicts $\Omega_B\,h_{50}^{2} \sim 0.04$, 
X--ray observations of dense clusters have indicated that $\Omega_B$
could be as large as $\leq 0.15$.  
Recent MACHO detections also suggest that there exist more baryons 
in our Galaxy than ever expected.  There is clearly a serious potential 
conflict between these observations and the theoretical prediction 
in the homogeneous big--bang model.  The situation is even crucial 
if high deuterium abundances, which were detected in Lyman--$\alpha$ 
absorption systems along the line of sight to high red--shifted quasars, 
are presumed to be primordial.
On the other hand, an {\em inhomogeneous big--bang model\/}, which allows inhomogeneous 
baryon density distribution, can predict $\Omega_B\,h_{50}^{2} \sim 0.1 - 0.2$.
Among the possible observable signatures of baryon
inhomogeneous cosmologies are the high abundances of heavier elements 
than lithium such as beryllium and boron\,\cite{kajino90}.
In an environment of baryon inhomogeneous distribution, neutrons can easily 
diffuse out of the fluctuations to form high density proton--rich 
and low density neutron--rich regions, where a lot of proton/neutron--rich 
radioactive isotopes can help produce the intermediate--to--heavy mass elements.

Another astrophysical site where the neutron--rich isotopes may play
a significant role in nucleosynthesis is the $\alpha$--process occurring
in supernovae. 
The nucleosynthesis in the high--entropy bubble is thought to proceed as
follows. Due to the high temperature, the previously produced nuclei up
to iron will be destroyed again by photodisintegration. At temperatures
of about 10$^{10}$\,K the nuclei would be dismantled into their
constituents, protons and neutrons. At slightly lower temperatures one is
still left with $\alpha$--particles. During the subsequent cooling of
the plasma the nucleons will recombine again, first to $\alpha$--particles,
then to heavier nuclei.
Depending on the exact temperatures, densities
and the neutron excess, quite different abundance distributions can be
produced in this {\em $\alpha$--rich freeze--out\/} (sometimes also called
{\em $\alpha$--process\/}). Temperature and density are
dropping quickly in the adiabatically expanding high--entropy bubble.
This will hinder the recombination of $\alpha$--particles into heavy nuclei,
leading in some scenarios to a high neutron density for an r--process,
at the end of the $\alpha$--process after freeze--out of charged particle reactions.

These astrophysical motivations have led us to critically study the
the role of radiative neutron capture reactions by neutron--rich Li-- 
and Be--isotopes theoretically in explosive nucleosynthesis.  
Since it is the focus in recent years to study the the nuclear reactions 
dynamics by the use of radioactive nuclear beams, our theoretical studies 
are also being tested experimentally.

\section{Calculation of Radiative--Capture Cross Sections}
Nuclear burning in explosive astrophysical environments produces
unstable nuclei which can again be targets for subsequent reactions. In
addition, it involves a
very large number of stable nuclei which are not yet fully explored
by experiments. Thus, it is necessary to be able to predict reaction
cross sections and thermonuclear rates with the aid of theoretical models.

In astrophysically relevant nuclear reactions two important reaction
mechanisms take place. These two mechanisms are compound--nucleus reactions
(CN) and direct reactions (DI).
The reaction mechanism and therefore
also the reaction model depends on the number of levels in the CN.
If one is considering only a few CN resonances the R--matrix theory
is appropriate.
In the case of a single resonance the R--matrix theory reduces to the
simple phenomenological Breit--Wigner
formula. If the level density of the CN is so high that there are 
many overlapping resonances,
the CN mechanism will dominate and the statistical
HF--model can be applied. Finally, if there are no CN resonances
in a certain energy interval the DI mechanism dominates and
one can use DI models, like Direct Capture (DC).

In the case of a single isolated resonance the resonant part 
of the cross section
is given by the well--known Breit--Wigner
formula\,\cite{Bre36,Bla62}:
\begin{equation}
\label{BW}
\sigma_{\rm r}(E) = 
\frac{\pi \hbar^2}{2 \mu E}
\frac{\left(2J+1\right)}{\left(2j_{\rm p}+1\right)\left(2j_{\rm t}+1\right)} 
\frac{\Gamma_{\rm in} \Gamma_{\rm out}}
{\left(E_{\rm r} - E\right)^2 + \frac{\Gamma_{\rm tot}^2}{4}} \quad ,
\end{equation}
where $J$ is the angular momentum quantum number and
$E_{\rm r}$ the resonance energy.
The partial widths of the entrance and exit channels
are $\Gamma_{\rm in}$  and $\Gamma_{\rm out}$, respectively.
The total width $\Gamma_{\rm tot}$
is the sum over the partial widths of all channels.
One important aspect is that
the particle width $\Gamma_{\rm p}$ can be related to
spectroscopic factors $S$ and the single--particle width $\Gamma_{\rm s.p.}$
by\,\cite{wie82,her95}
\begin{equation}
\label{SF}
\Gamma_{\rm p} = C^2 S \Gamma_{\rm s.p.} \quad,
\end{equation}
where $C$ is the isospin Clebsch--Gordan coefficient.
The single--particle width $\Gamma_{\rm s.p.}$ can be calculated
from the scattering phase shifts of a scattering potential with
the potential depth determined by matching the resonance energy.

The nonresonant part of the cross section can be obtained
using the DC model\,\cite{kim87,obe91,moh93}:
\begin{equation}
\label{NR}
\sigma^{\rm nr} = \sum_{c} \: C^{2} S_c\sigma^{\rm DC}_c \quad .
\end{equation}
The sum extends over all bound states   
in the final nuclei. The DC cross sections $\sigma^{\rm DC}_c$ are
essentially
determined by the overlap of the scattering wave function
in the entrance channel, the bound--state wave function
in the exit channel and the multipole transition--operator.

The total cross section can be calculated by summing
over the resonant (Eq.~\ref{BW}) and nonresonant parts
(Eq.~\ref{NR}) of the cross section (if the
widths of the resonances are broad, also an
interference term has to be added).
For both parts the spectroscopic factors have to be known. They can be obtained
from other reactions, e.g., the spectroscopic factors
necessary for calculating
A(n,$\gamma$)B can be extracted
from the reaction A(d,p)B. The $\gamma$--widths can be
extracted from reduced electromagnetic transition
strengths. 
For unstable nuclei 
where only limited or even no
experimental information is available, the
spectroscopic factors and electromagnetic transition strengths 
can also be extracted from nuclear structure models like the shell model (SM).

The most important ingredients in the potential models are the wave functions 
for the scattering and bound states in the entrance and exit channels.
This is the case for the DC cross sections $\sigma^{\rm DC}_c$ in
Eq.~\ref{NR} as well as for the calculation of the single--particle width
$\Gamma_i$ in Eq.~\ref{SF}.
For the calculation of these wave functions we use real folding potentials
which are given by\,\cite{obe91,kob84}
\begin{equation}
\label{FO}
V(R) = 
  \lambda\,V_{\rm F}(R) 
  = 
  \lambda\,\int\int \rho_a({\bf r}_1)\rho_A({\bf r}_2)\,
  v_{\rm eff}\,(E,\rho_a,\rho_A,s)\,{\rm d}{\bf r}_1{\rm d}{\bf r}_2 \quad ,
\end{equation}
with $\lambda$ being a potential strength parameter close
to unity, and $s = |{\bf R} + {\bf r}_2 - {\bf r}_1|$,
where $R$ is the separation of the centers of mass of the
projectile and the target nucleus.
The density can been derived from measured
charge distributions\,\cite{vri87} or from nuclear structure models (e.g.,
Hartree--Fock calculations) and the effective nucleon--nucleon
interaction $v_{\rm eff}$
has been taken in the DDM3Y parametrization\,\cite{kob84}.
The imaginary part of the potential 
is very small because of the small flux into other reaction channels
and can be neglected in most cases involving neutron capture
by neutron--rich target nuclei.

\section{Reaction Rates for Li-- and Be--Isotopes}

The parameters for the resonant and nonresonant contributions to the
reaction rates are listed in Tables \ref{res} and \ref{nres},
respectively. In the tables we give experimental values if
available. Otherwise the excitation energies, spectroscopic
factors, neutron-- and $\gamma$--widths were calculated with the
shell model. We used the code OXBASH\,\cite{bro84} for the calculations.
For normal parity states we employed the interaction
(8--16)POT of Cohen and Kurath\,\cite{coh65}.
For nonnormal parity states we used the WBN interaction of
Warburton and Brown\,\cite{war92}.

With Eq.~\ref{BW} the resonant reaction rate can be derived as

\begin{eqnarray}
\label{resrate}
N_{\rm A}\left\langle\sigma v\right\rangle_{\rm r} & = & 1.54
\times 10^5 \mu^{-3/2} T_9^{-3/2}\\\nonumber
&& \sum_i {\omega \gamma_i \exp(-11.605 E_{\rm r} / T_9)\,{\rm cm}^3
\,{\rm mole}^{-1}\,{\rm s}^{-1}} \quad ,
\end{eqnarray}
where $T_9$ is the temperature in $10^9$K, $E_r$ the resonance energy
in the c.m.~system (in MeV), and the resonance strength $\omega \gamma$ (in eV) is
given by

\begin{equation}
\omega \gamma = \frac{2J+1}{(2j_{\rm p}+1)(2j{\rm _t}+1)} \frac{\Gamma_{\rm in}
\Gamma_{\rm out}}{\Gamma_{\rm tot}} \quad .
\end{equation}
The partial widths
of the entrance and exit channel, $\Gamma_{\rm in}$ and $\Gamma_{\rm out}$,
are in the case of (n,$\gamma$)--reactions the neutron-- and
$\gamma$--widths. Since the neutron width is usually much larger than
the $\gamma$--width, the total width $\Gamma_{\rm tot}$ is practically
identical with the neutron--width.

In Table \ref{res} we list the excitation energies, resonance energies,
neutron-- and $\gamma$--widths and the resonance strengths of the
resonances.

The nonresonant capture cross section is parametrized as 
\begin{equation}
\sigma_{\rm nr} (E)=
A / \sqrt{E} + B \sqrt{E} -C  E^D\: , 
\end{equation}
with $[A]={\rm \mu b\, MeV^{1/2}}$,
$[B]={\rm \mu b \, MeV^{-1/2}}$, and $[C]={\rm \mu b\, MeV^{-D}}$.
The parameters $A, B, C$ and $D$ are listed in Table \ref{nres}.
Using this equation, we obtain for the reaction rate
\begin{eqnarray}
\label{nresrate}
N_{\rm A}\left\langle\sigma v\right\rangle_{\rm nr} & = & \bigg(
836.565 A\mu^{-1/2}+108.130 B \mu^{-1/2} T_9 \nonumber \\
& &  -277.097 C\mu^{-1/2}
\frac{\Gamma(2+D)}{\left.11.605^D\right.}    
T_9^{D+1/2}\bigg)\,{\rm cm^3\, s^{-1}\,mole^{-1}}\; ,
\end{eqnarray}
where $\mu$ ist the reduced mass in units of the atomic mass unit
and $\Gamma(z)$
is the Euler gamma function.

The total reaction rate is given as the sum of the resonant (Eq.~\ref{resrate} 
and nonresonant (Eq.~\ref{nresrate}) part.

\begin{table}
\caption[RES]{\label{res} Resonance parameters}
\begin{tabular}{|lrrrrrr|}
\hline
Reaction &
\multicolumn{1}{c}{$E_{\rm x}$} &
\multicolumn{1}{c}{$E_{\rm n}$} &
\multicolumn{1}{c}{$J^{\pi}$} &
\multicolumn{1}{c}{$\Gamma_{\rm n}$} &
\multicolumn{1}{c}{$\Gamma_{\gamma}$} &
\multicolumn{1}{c|}{$\omega\gamma$} \\
& \multicolumn{1}{c}{(MeV)} &
\multicolumn{1}{c}{(MeV)} &&
\multicolumn{1}{c}{(eV)} &
\multicolumn{1}{c}{(eV)} &
\multicolumn{1}{c|}{(eV)} \\
\hline
$^7$Li(n,$\gamma$)$^8$Li & 2.26 & 0.227 & $3^+$ & $3.1 \times 10^4$ &
$0.07$ & $0.061$ \\
$^8$Li(n,$\gamma$)$^9$Li & 4.31 & 0.247 & $5/2^-$ & $1 \times 10^5$ &
0.11 & $0.066$ \\
$^9$Be(n,$\gamma$)$^{10}$Be & 7.371 & 0.559 & $3^-$ & $1.57 \times 10^4$ &
0.661 & 0.578 \\
& 7.542 & 0.73 & $2^+$ & $6.3 \times 10^3$ &
0.814 & 0.509 \\
\hline 
\end{tabular}
\end{table}

\begin{table}[htb]
\caption[di cross section]{\label{nres}Parametrization of the nonresonant cross
section (see text).} 
\begin{center}
\begin{tabular}{|llrrrrr|}\hline
 & \multicolumn{1}{c}{$A$}& \multicolumn{1}{c}{$B$} & 
  \multicolumn{1}{c}{$C$} & \multicolumn{1}{c}{$D$} 
& \multicolumn{2}{c|}{$\sigma_{\rm nr}({\rm\mu b})\:\:{\rm at}\:\:{\rm 30\, keV}$} 
\\ \hline
& & & & & \multicolumn{1}{c}{This} 
 & \multicolumn{1}{c|}{Rauscher}
 \\ 
 & & & & & 
 \multicolumn{1}{c}{work} &
 \multicolumn{1}{c|}{{\em et al.}\cite{rau94}} \\
\hline
${\rm ^{7}Li(n,\gamma)^{8}Li}$ & $6.755^{\rm a}$ 
& \multicolumn{1}{c}{---} & \multicolumn{1}{c}{---} 
& \multicolumn{1}{c}{---} & \multicolumn{1}{c}{$39.000$} 
& \multicolumn{1}{c|}{---}  \\
${\rm ^{8}Li(n,\gamma)^{9}Li}$ & $2.909$ & \multicolumn{1}{c}{---} 
& \multicolumn{1}{c}{---} &\multicolumn{1}{c}{---}
& \multicolumn{1}{c}{$16.795$} & $30.392$
 \\
${\rm ^{9}Be(n,\gamma)^{10}Be}$ &  $1.147^{\rm a}$ & $11.000$ 
& $6.815$ & $0.962$ 
& $8.294$ & $6.622$ \\
${\rm ^{10}Be(n,\gamma)^{11}Be}$ & $0.132$ & $24.000$ & $15.725$ & $0.914$ 
& $4.281$ & $3.943$ \\ 
${\rm ^{11}Be(n,\gamma)^{12}Be}$ & \multicolumn{1}{c}{---} 
& $7.000$ & $4.851$ & $0.887$ 
& $0.996$ & $2.373$ \\ \hline
\end{tabular}
\end{center}
\small
$^{\rm a}$extracted from experimental thermal cross section\,\cite{sea92} 
\end{table}

\subsection{$^7$Li(n,$\gamma$)$^8$Li}

The cross section of the reaction $^7$Li(n,$\gamma$)$^8$Li is
well known (see, e.g.,\,\cite{bla96}). The cross section is
dominated by s--wave capture to the $^8$Li ground state and
a resonance at 227\,keV neutron energy. Using the spectroscopic
factors of Cohen and Kurath\,\cite{coh67} yields a thermal
cross section of $8.2 \times 10^{-2}$\,b, which is a factor
1.8 higher than the experimental value of $4.54 \times 10^{-2}$\,b.
The shell model calculation is purely p--shell and does not
include excitations to other oscillator shells. Therefore the
spectroscopic amplitude of 0.977 for a p$_{3/2}$--transition to
the ground state of $^8$Li might be too high.

For the resonance, however, we find excellent agreement between
calculation and experiment. The calculated width --- using the
folding potential and spectroscopic amplitudes from
Cohen and Kurath\,\cite{coh67}
--- is 28.9\,keV, almost identical to the known value of $31 \pm 7$\,keV.

\subsection{$^8$Li(n,$\gamma$)$^9$Li}

The resonance at 247 keV is a $5/2^-$ state\,\cite{van84}.
With a total width of 100\,keV the resonance strength
is determined by the $\gamma$--width which was previously
estimated with 0.56\,eV\,\cite{mal88}. A shell model calculation
yielded a width $\Gamma_{\gamma} = 0.11$\,eV. Therefore the
resonance strength is a factor 5 smaller than previously
assumed\,\cite{rau94}.

The calculated thermal cross section, resulting from s--wave
capture to the ground state and first excited state in $^9$Li,
is 1.94 $\times 10^{-2}$\,b and is smaller than the value
of $3.51 \times 10^{-2}$\,b given by Rauscher {\em et al.}\,\cite{rau94}.

\subsection{$^9$Be(n,$\gamma$)$^{10}$Be}

Like in the reaction $^7$Li(n,$\gamma$)$^8$Li the spectroscopic
factors of Cohen and Kurath\,\cite{coh67} are a little too high.
The thermal cross section is dominated by the transition to the
$^{10}$Be ground state with a theoretical spectroscopic factor
of 2.36. With this value the calculated thermal cross section
is $1.06 \times 10^{-2}$\,b, compared to experimental cross section
of $7.6 \times 10^{-3}$\,b. With the spectroscopic factor given by
Mughabghab\,\cite{mug85} of 1.45 the calculated cross section
would be close to the experimental value. For high temperatures the
p--wave capture to excited states has to be taken into account.

Two resonances are known at 559\,keV and 730\,keV. The total widths
are known experimentally. We have calculated the $\gamma$--widths
which were only estimated previously. Both resonance strengths are larger
than the previous estimates, for the 559\,keV resonance the
enhancement is one order of magnitude.

With the higher resonance strengths and the p--wave contribution the
reaction rate is clearly higher compared to Ref.~\cite{rau94}.

\subsection{$^{10}$Be(n,$\gamma$)$^{11}$Be}

Cross section and reaction rate of this reaction were recently
determined experimentally with the help of the inverse
Coulomb dissociation\,\cite{men96}. They supported their experimental
values by a direct capture calculation. In order to reproduce the
experimental data they enhanced the spectroscopic factors to the
$^{11}$Be ground state by 20\%. Our calculation confirms the results.
Using the spectroscopic factors from the (d,p)--reaction\,\cite{ajz90}
the calculated cross section is a little smaller than the experimental.
The results are grossly different from the rate given in Ref.~\cite{rau94}.

\subsection{$^{11}$Be(n,$\gamma$)$^{12}$Be}

There is no resonant contribution to the reaction rate. The transition
is a p--wave capture from the $1/2^+$ ground state of $^{11}$Be 
to the ground state of $^{12}$Be and the $0^+$ state at 2.7\,MeV
excitation energy, while the transition to the $2^+$ state at 2.1\,MeV
is negligible.

\section{Discussion}

The new reaction rates could change the reaction flow in the inhomogeneous big
bang nucleosynthesis. The smaller rate for $^8$Li(n,$\gamma$)$^9$Li could mean
that the main reaction flow will proceed through the reaction $^8$Li($\alpha$,n)$^{11}$B.
The higher rate for $^9$Be(n,$\gamma$)$^{10}$Be might give more importance to this
reaction. Detailed network calculations with the new rates are planned for the
near future.

\section*{Acknowledgments}
We thank the  Fonds zur F\"orderung der
wissenschaftlichen Forschung in \"Osterr\-reich (project S7307--AST)
and the \"Ostereichische Nationalbank (project 5054)
for their support.


\section*{References}
\begin{thebibliography}{99}
 
\bibitem{kajino90}  T. Kajino and R.N. Boyd,
{\em Astrophys. J.} {\bf 359}, 267 (1990).

\bibitem{Bre36}
R.G. Breit and E.P. Wigner, {\em Phys. Rev.} {\bf 49}, 519 (1936).
 
\bibitem{Bla62}
J.M. Blatt and V.F. Weisskopf, {\it Theoretical Nuclear
Physics}, (Wiley \& Sons, New York, 1962).
 
\bibitem{wie82}
M. Wiescher and K.-U. Kettner,
{\em Astrophys. J.} {\bf 263}, 891 (1982).
 
\bibitem{her95}
H. Herndl, J. G\"orres, M. Wiescher, B.A. Brown and L. van Wormer,
{\em Phys. Rev.} C {\bf 52}, 1078 (1995).
 
\bibitem{kim87} K.H. Kim, M.H. Park and B.T. Kim, 
{\em Phys. Rev.} C {\bf 35}, 363 (1987).
 
\bibitem{obe91}
H. Oberhummer and G. Staudt, in {\it Nuclei in the Cosmos},
ed. H. Oberhummer,
(Springer--Verlag, Berlin, New York, 1991) p. 29.
 
\bibitem{moh93} P. Mohr, H. Abele, R. Zwiebel, G. Staudt, 
H. Krauss, H. Oberhummer, A. Denker, J.W. Hammer and G. Wolf,
{\em Phys. Rev.} C {\bf 48}, 1420 (1993).
 
\bibitem{kob84} A.M. Kobos, B.A. Brown, R. Lindsay and G.R. Satchler, 
{\em Nucl. Phys.} A {\bf 425}, 205 (1984).
 
\bibitem{vri87} H. de Vries, C.W. de Jager and C. de Vries, 
{\em At. Data Nucl. Data Tables} {\bf 36}, 495 (1987).

\bibitem{bro84} B.A. Brown, A. Etchegoyen, W.D.M. Rae and
N.S. Godwin, code OXBASH, 1984 (unpublished).

\bibitem{coh65} S. Cohen and D. Kurath, {\em Nucl. Phys.} {\bf 73},
1 (1965).

\bibitem{war92} E.K. Warburton and B.A. Brown, {\em Phys. Rev.} C {\bf 46},
923 (1992).

\bibitem{bla96} J.C. Blackmon, A.E. Champagne, J.K. Dickens, J.A. Harvey,
M.A. Hofstee, S. Kopecky, D.C. Larson, D.C. Powell, S. Raman and
M.S. Smith, {\em Phys. Rev.} C {\bf 54}, 383 (1996).

\bibitem{coh67} S. Cohen and D. Kurath, {\em Nucl. Phys. A} {\bf 101},
1 (1967).

\bibitem{sea92} V.F. Sears, {\em Neutron News}, Vol. 3, 26 (1992)

\bibitem{van84} A.G.M. van Hees and P.M.W. Glaudemans, {\em Z. Phys.} A
{\bf 315}, 223 (1984).

\bibitem{mal88} R.A. Malaney and W.A. Fowler, {\em Ap. J.} 333, 14
(1988).

\bibitem{rau94} T. Rauscher, J.H. Applegate, J.J. Cowan, F.-K.
Thielemann and M. Wiescher, {\em Ap. J.} 429, 499 (1994).

\bibitem{mug85} S.F. Mughabghab, {\em Phys. Rev. Lett.} {\bf 54}, 986 (1985).

\bibitem{men96} A. Mengoni, T. Otsuka, T. Nakamura and M. Ishihara,
{\em Nucl. Phys.} A (Proc. Suppl.), in print.

\bibitem{ajz90} F. Ajzenberg--Selove, {\em Nucl. Phys.} A {\bf 506}, 1 (1990).
\end{thebibliography}
\end{document}